# Resonant Raman and photoluminescence spectra of suspended molybdenum disulfide

*Jae-Ung Lee[1], Kangwon Kim[1] and Hyeonsik Cheong*

Department of Physics, Sogang University, Seoul 121-742, Korea

**E-mail:** hcheong@sogang.ac.kr

**Notes:** [1]These authors contributed equally.

**Abstract**

Raman and photoluminescence (PL) measurements were performed on suspended and supported $MoS_2$ up to 6 trilayers (TLs). The difference in the dielectric environment of suspended and supported $MoS_2$ does not lead to large differences in the PL or Raman spectra. Most of the apparent difference in the PL intensity can be explained by the interference effect except for 1TL $MoS_2$. The positions of the Raman peaks are not much different between suspended and supported $MoS_2$. The relative intensities, on the other hand, show some differences between suspended and supported samples. In particular, the inter-layer vibration modes are enhanced near resonance with C excitons, with the breathing modes becoming relatively stronger. Up to 4 breathing modes are resolved in suspended 6TL $MoS_2$.

## 1. Introduction

Two-dimensional (2D) semiconductors are drawing much interest as possible candidates for flexible electronics system [1]. Molybdenum disulfide ($MoS_2$) is a prototypical 2D semiconducting transition metal dichalcogenide (TMD) and has been most extensively studied. Single-layer $MoS_2$ is composed of two S layers and one Mo layer arranged in hexagonal lattices [2]. These layers are covalently bonded to one another and form a 'trilayer' (TL). Thicker $MoS_2$ is formed by stacking such TLs, and the most common stacking type is the so-called 2H stacking order [3]. In few-layer or bulk $MoS_2$, the TLs are bonded to each other through weak van der Waals interaction. Single-layer (1TL) $MoS_2$ has a direct bandgap whereas thicker $MoS_2$ has an indirect gap. The studies on the optical properties of $MoS_2$ have yielded many interesting results. Strong photoluminescence (PL) has been observed for few-layer $MoS_2$. Whereas 1TL $MoS_2$ shows a strong PL signal due to the direct bandgap, the PL signal due to the indirect bandgap in thicker $MoS_2$ can be used to estimate the thickness as the peak position depends on the number of layers [4,5]. Several exciton states with large exciton binding energies due to reduced dielelctric screening [6–9] and the control of valley polarization by using circularly polarized light [10–12] have been reported.

Raman spectroscopy is a most widely used tool in the study of 2D crystals. For $MoS_2$, the number of layers can be identified by the separation of the main optical phonon modes [13,14] or from the low-frequency inter-layer vibrations [14–16]. Also, interesting resonance Raman behaviors have been reported [14,17–22]. When the excitation energy matches with the A (or B) exciton energy at ~1.85 eV (~2.05 eV), several new Raman features appear due to the resonance. Furthermore, selective enhancement of some peaks due to symmetry dependent exciton-phonon interactions [23,24] has been observed near the resonance with the C exciton state. Similar resonance effects are also seen in other TMD materials [25–28]. However, the details of the mechanism through which the excitonic resonance affects the Raman spectrum are not understood well.

Since most Raman measurements are performed on samples placed on substrates, the influence of the substrate adds complications. There have been few publications on the substrate effects. Some groups [29,30] tried to understand the substrate effects by measuring Raman or PL spectra of $MoS_2$ samples on different substrates. Scheuschner *et al.* [31] measured PL and Raman spectra of $MoS_2$ samples suspended over a circular hole in the substrate. However, none of these previous studies paid attention to the resonance Raman effect which is expected to be most affected by the substrate. In this work, we measured the Raman and PL spectra of suspended $MoS_2$ with thickness from 1TL to 6TL. In order to study the resonance Raman effect, we used 3 excitation energies: 1.96, 2.41, and 2.81 eV, which correspond to resonance with the A/B excitons, non-resonant, and resonance with the C exciton, respectively. By comparing the Raman and PL spectra of suspended samples with those of supported ones, we isolated the effects of the substrate.

## 2. Experimental

Si substrates with a 300-nm $SiO_2$ top layer, patterned with circular holes (2~7 μm in diameter and 3 μm in depth) by photolithography and dry etching, were used. The $MoS_2$ samples were prepared directly on the pre-patterned $SiO_2$/Si substrates by mechanical exfoliation from single-crystal bulk $MoS_2$ flakes (SPI supplies). To avoid possible sample-to-sample variation, the comparison between suspended and supported regions is performed on the same sample. Figure 1(a) is a schematic of the sample. Representative optical microscope images of the samples are shown in Fig. 1(b). The number of TLs was determined by combining Raman and PL measurements [4,13]. The 441.6-nm (2.81 eV) line of a He-Cd laser, 514.5-nm (2.41 eV) line of an Ar ion laser, and the 632.8-nm (1.96 eV) line of a He-Ne laser are used as excitation sources. The laser beam was focused onto a sample by a 50× microscope objective lens (0.8 N.A.), and the scattered light was collected and collimated by the same objective. The scattered signal was dispersed by a Jobin-Yvon Horiba iHR550 spectrometer (2400 grooves/mm) and detected with a liquid-nitrogen-cooled back-illuminated charge-coupled-device (CCD) detector. To access the low-frequency range below 100 $cm^{-1}$, volume holographic filters (Ondax for 514.5-nm excitation, OptiGrate for 441.6 and 632.8-nm excitations) were used to clean the laser lines and reject the Rayleigh-scattered light. The laser power was kept below 50 μW in order to avoid local heating. We observed significant damages on suspended samples when the laser power is higher than 200 μW due to a low thermal conductivity of $MoS_2$ [32].

## 3. Results

Figure 2(a) shows PL spectra of suspended and supported $MoS_2$ samples with thickness from 1TL to 6TL. The peaks at ~1.85 eV and ~2.05 eV are assigned to A (A⁻) and B excitons, respectively, associated with the direct gap at the K (K′) point. The peak at 1.2~1.5 eV appears only in thicker samples and are ascribed to the PL due to the indirect gap [4,33]. For the same thickness, the PL intensity tends to be stronger for the suspended sample. The most dramatic difference is seen for the 1TL sample, where the PL from the suspended part of the sample is about 40 times stronger than that from the supported part. Also, the indirect PL of the 2TL sample is much stronger in the suspended part. Similar effects have been reported before [4,31] and ascribed to the suppression of A excitons due to doping in supported $MoS_2$ [31,34].

The intensity of the PL signal can be affected by the interference effect due to multiple reflection in the substrate [35,36]. The enhancement factors due to interference were calculated by considering multiple reflections at the air/$MoS_2$, $MoS_2$/$SiO_2$, and $SiO_2$/Si interfaces. The interference effect on the excitation laser as well as the PL emission signal was included. Figure 2(b) compares the enhancement factors of suspended and supported $MoS_2$ for 1TL and 2TL. Note that suspended $MoS_2$ also shows some interference effect due to the thin-film interference inside the $MoS_2$ layer. The refractive indices of $MoS_2$ is adopted from the literature values for 1TL $MoS_2$ [37]. The energy range of the calculation is limited by the available data for the refractive indices. By dividing the measured PL intensity by the calculated enhancement factor, one can remove the effect of the interference. Figure 2(c) compares the PL spectra of suspended and supported $MoS_2$ after the interference effect is removed. Again, the energy range of the corrected spectra is limited by the available refractive indices data. It is evident that the large difference between the intensities of the suspended and supported 1TL MoS$_2$ cannot be explained by the interference effect and the effect of doping on the supported MoS$_2$ [31,34] may still play a major role. For 2TL, however, most of the difference is removed by considering the interference effect. Therefore, the effect of doping, if any, should be relatively small for 2TL or thicker MoS$_2$. Another possible mechanism for reduced PL from suspended MoS$_2$ is the loss of photo-excited carriers through the substrate. This additional quenching channel of the photo-excited carriers would also be most effective for 1TL. We also notice that the indirect PL is stronger than the direct PL after the interference effect is removed. For suspended samples, small fringes are observed in the PL spectra, which is due the interference between the sample and the bottom of the hole [31]. The energy spacing of ~250 meV corresponds to a distance of ~2.5 μm, which is similar to the depth of the holes ~3 μm.

Now we turn to Raman measurements of suspended few-layer MoS$_2$. As reported by several groups [17,20,21,38], MoS$_2$ exhibits anomalous resonance Raman behaviors. We used 3 different excitation energies (1.96, 2.41, and 2.81 eV) in order to study the resonance effects. In Fig. 3, the 2.41 eV excitation can be regarded as non-resonant, and only the first-order E$_{2g}^1$ and A$_{1g}$ Raman modes are observed. These modes are often used to determine the number of layers of MoS$_2$ [13]. For few-layer samples, the layer-by-layer vibrations, so called shear and breathing modes [15,16] are also observed as shown in Fig. 3 (b). The 1.96 eV excitation energy matches the A or B exciton states, which leads to anomalous resonant effects. The central peak, the broad peak centered at the Rayleigh scattered laser line, appears due to the resonance excitation of exciton states mediated by acoustic phonon scattering [17]. In addition, many second order Raman peaks are enhanced due to resonance with exciton or exciton-polariton states [17]. For the suspended 1TL sample, the second-order peaks as well as the central peak are somewhat sharper than those of the supported sample. In contrast, the differences are almost negligible for 2TL. The excitation energy of 2.81 eV is close to the C exciton state in $MoS_2$[6,9,23]. The selective enhancement of the $E_{2g}^1$ mode with respect to the $A_{1g}$ mode has been attributed to symmetry dependent exciton-phonon coupling. [23] The enhancement of forbidden Raman modes, $E_{1g}$ and $A_{2u}$, are also reported [17,24]. The difference in the overall intensities of supported and suspended samples can be explained by the interference effect. The $E_{2g}^1/A_{1g}$ intensity ratio shows difference between suspended and supported samples, especially for 2TL. Because this ratio is sensitive to the exciton-phonon coupling, small changes in the exciton states due to the (absence of) supporting substrate would be responsible for such differences.

Figure 4 compares the Raman spectra of suspended and supported $MoS_2$ for 1TL – 6TL in the low-frequency region and near the main modes. For off-resonant excitation (2.41 eV), the main $E_{2g}^1$ and $A_{1g}$ modes (Fig. 4d) are very similar between suspended and supported samples for all thickness. Upon close inspection, one can see that the peak positions are slightly different. For 1TL, the Raman peaks of the suspended samples are slightly blueshifted with respect to those of the supported samples. For 2TL, the peaks almost coincide. For 3TL or thicker, the Raman peaks of the suspended samples are slightly redshifted. Such small shifts can be caused by a number of reasons: doping due to the substrate, residual strain or interaction with the substrate. We note that there has been a report that the lattice constant of $MoS_2$ is modified by the interaction with the substrate[39], which should depend on the thickness of $MoS_2$. The low-frequency modes (Fig. 4c) are virtually identical between suspended and supported samples. Since the shear and breathing modes are due to rigid vibrations of the TLs, small differences in doping or strain should not affect them.

On the other hand, for resonant excitations (1.96 and 2.81 eV), there are clear differences in the peak positions or shapes between suspended and supported samples. For the 1.96 eV excitation, the high-frequency range of the Raman spectrum shows many strong second-order peaks activated by resonance with exciton states. [17,38,40] For example, the shoulder-like peaks near the $E_{2g}^1$ and $A_{1g}$ modes have been ascribed as being due to finite-momentum phonons of the $E_{2g}^1$ and $A_{1g}$ branch activated by resonance with exciton-polaritons. [17] Since the exciton states in $MoS_2$ are influenced by the presence of the substrate, the resonance-induced peaks should be affected. On the other hand, the low-frequency region does not show a big difference except for the difference in the central peak of 1TL as described already in Fig. 3. For the 2.81 eV excitation, the breathing modes are relatively enhanced for suspended as well as supported samples. Up to 4 breathing modes are observed in 6TL, as opposed to only 2 breathing modes observed for the 2.41 eV excitation. Since 2.81 eV is close to the C exciton energy, this enhancement of the breathing modes indicates that the exciton-phonon interaction is selectively enhanced for the breathing modes. Because the C exciton state is derived from the $d_{z^2}$ orbital of Mo and the $p_{x,y}$ orbitals of S atoms whereas A or B exciton states are mostly derived from the $d_{z^2}$ orbital of Mo, it is no surprise that the inter-layer vibrations are enhanced more for the resonance with the C excitons. However, in the case of main modes, the $E_{2g}^1$ in-plane vibration mode is selectively enhanced near resonance with C excitons, whereas the vibrations of the breathing modes are orthogonal to the $E_{2g}^1$ mode. It is clear that the enhancement of the exciton-phonon interactions cannot be explained simply in terms of the symmetry of the exciton states and the direction of the vibrations. Further studies are needed to understand the details of the exciton-phonon interactions in $MoS_2$. Figure 4e also shows that the breathing modes are more clearly resolved for suspended samples. This is reasonable, because breathing modes tend to be damped by the interaction with the substrate. For the main modes in the high frequency range, the comparison between suspended and supported samples is similar to the case of the 2.41 eV excitation except for the difference in the $E_{2g}^1/A_{1g}$ intensity ratio as described in Fig. 3.

The overall similarity between the Raman spectra of suspended and supported $MoS_2$ is intriguing. In a simple picture, the large exciton binding energy is a result of reduced dielectric screening in 2-dimensional materials. When the dielectric environment is modified, the binding energy should be sensitively affected. For example, the binding energy of $MoS_2$ on a substrate should be smaller than that of suspended $MoS_2$. On the other hand, it has been suggested that the reduction of the binding energy is balanced by renormalization of the bandgap energy and the exciton energy is not significantly different between suspended and supported $MoS_2$ [8]. This is supported by the small difference in the PL energy between suspended and supported $MoS_2$. Our results indicate that the difference in the exciton binding energy, which means difference in the exciton wave functions, does not affect exciton-phonon interactions except for small differences. There is no simple explanation for this weak correlation between the dielectric environment and the exciton-phonon interaction. More theoretical work is needed to understand the interplay between excitons and phonon modes in $MoS_2$ and TMDs.

## 4. Conclusion

In conclusion, we performed Raman and PL measurements on suspended and supported $MoS_2$ up to 6TL. We showed that most of the apparent difference in the PL intensity can be explained by the interference effect, but the intensity of the A exciton peak of suspended 1TL $MoS_2$ is clearly enhanced due to reduced doping. From Raman measurements with 3 different excitation energies, the effects of resonance with exciton or exciton-polariton states are compared for suspended and supported $MoS_2$. For off-resonance excitation (2.41 eV) the main phonon peaks show small relative shifts due to interaction with the substrate. For resonance with A or B excitons (1.96 eV), the central peak of the low-frequency region tends to be sharper for suspended samples, and the resonance-activated higher order peaks show small differences between suspended and supported samples. For resonance with C excitons (2.81 eV), the inter-layer vibration modes are enhanced, with the breathing modes becoming relatively stronger. In suspended 6TL $MoS_2$, up to 4 breathing modes are resolved. The observed weak correlation between the dielectric environment and Raman spectra of $MoS_2$ is intriguing and calls for more theoretical studies.

**Acknowledgements**

This work was supported by the National Research Foundation (NRF) grants funded by the Korean government (MSIP) (Nos. 2011-0013461 and 2011-0017605) and by a grant (No. 2011-0031630) from the Center for Advanced Soft Electronics under the Global Frontier Research Program of MSIP.

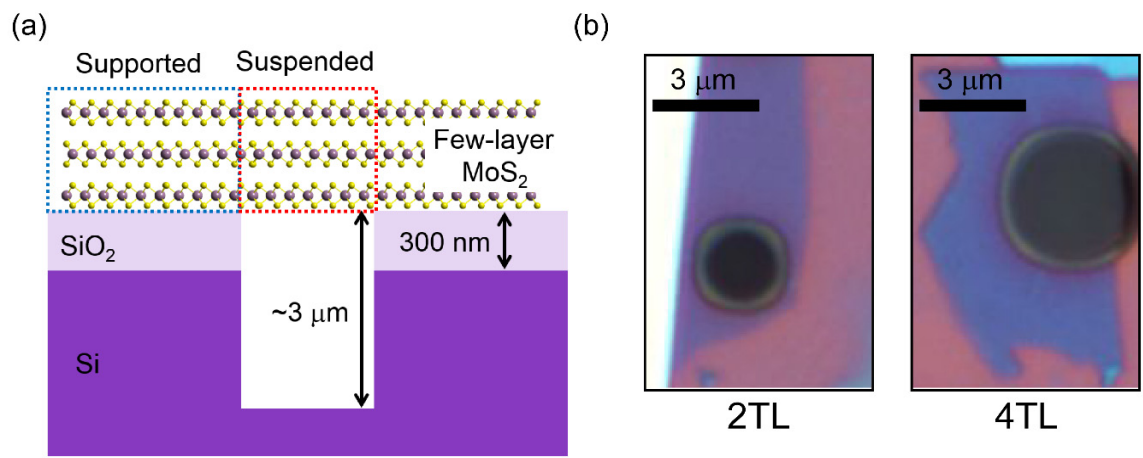

**Figure 1** (a) Schematic of suspended MoS$_2$ sample. (b) Optical microscope images of suspended 2TL and 4TL MoS$_2$.

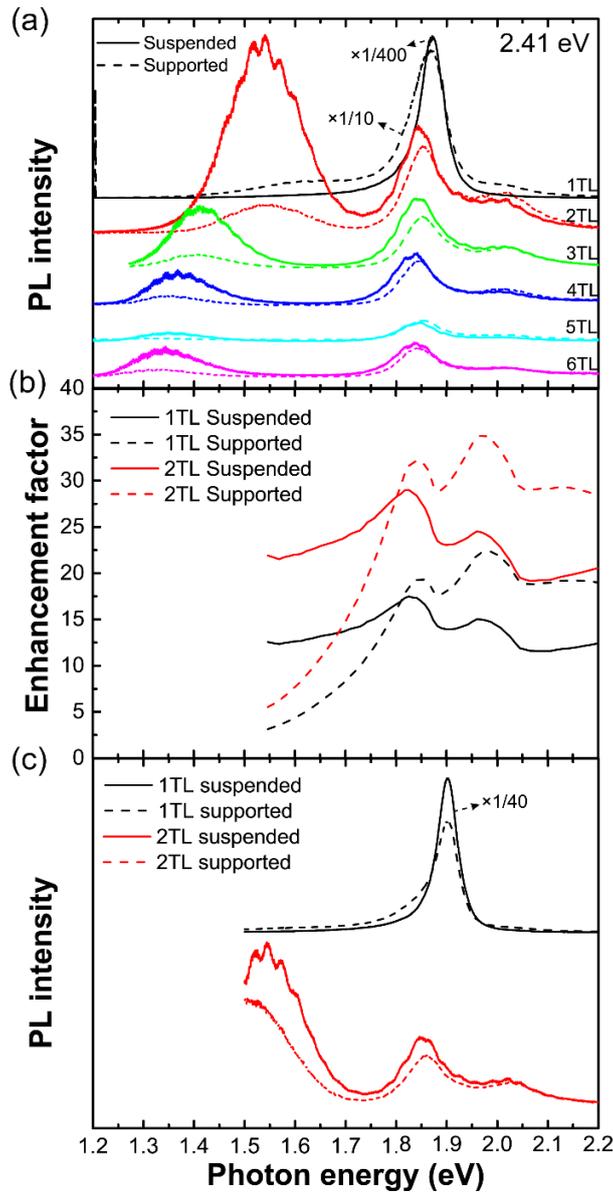

**Figure 2** (a) PL spectra of suspended (solid lines) and supported (dashed lines) MoS₂ with thickness from 1TL to 6TL measured with 2.41 eV (514.5 nm) excitation. (b) Enhancement factor due to interference effect for 2.41 eV (514.5 nm) excitation. (c) PL spectra of suspended (solid line) and supported (dashed line) 1TL and 2TL MoS₂ with the interference effect removed.

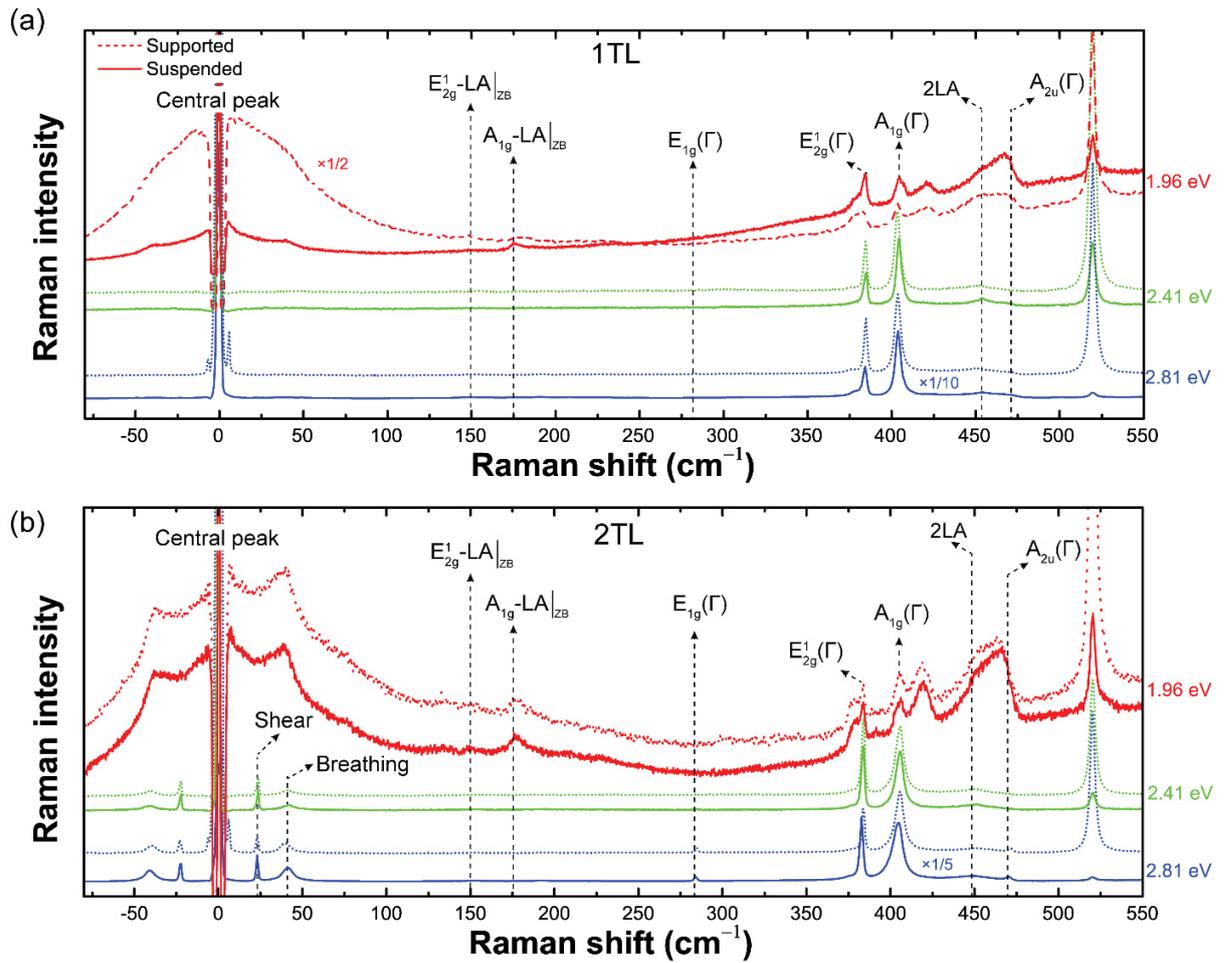

**Figure 3** Raman spectra of suspended (solid lines) and supported (dashed lines) (a) 1TL and (b) 2TL MoS₂ with 1.96 (red), 2.41 (green), and 2.81 eV (blue) excitation energies. The spectral intensities for different excitation energies are normalized by the intensity of the Si peak from bare SiO₂/Si.

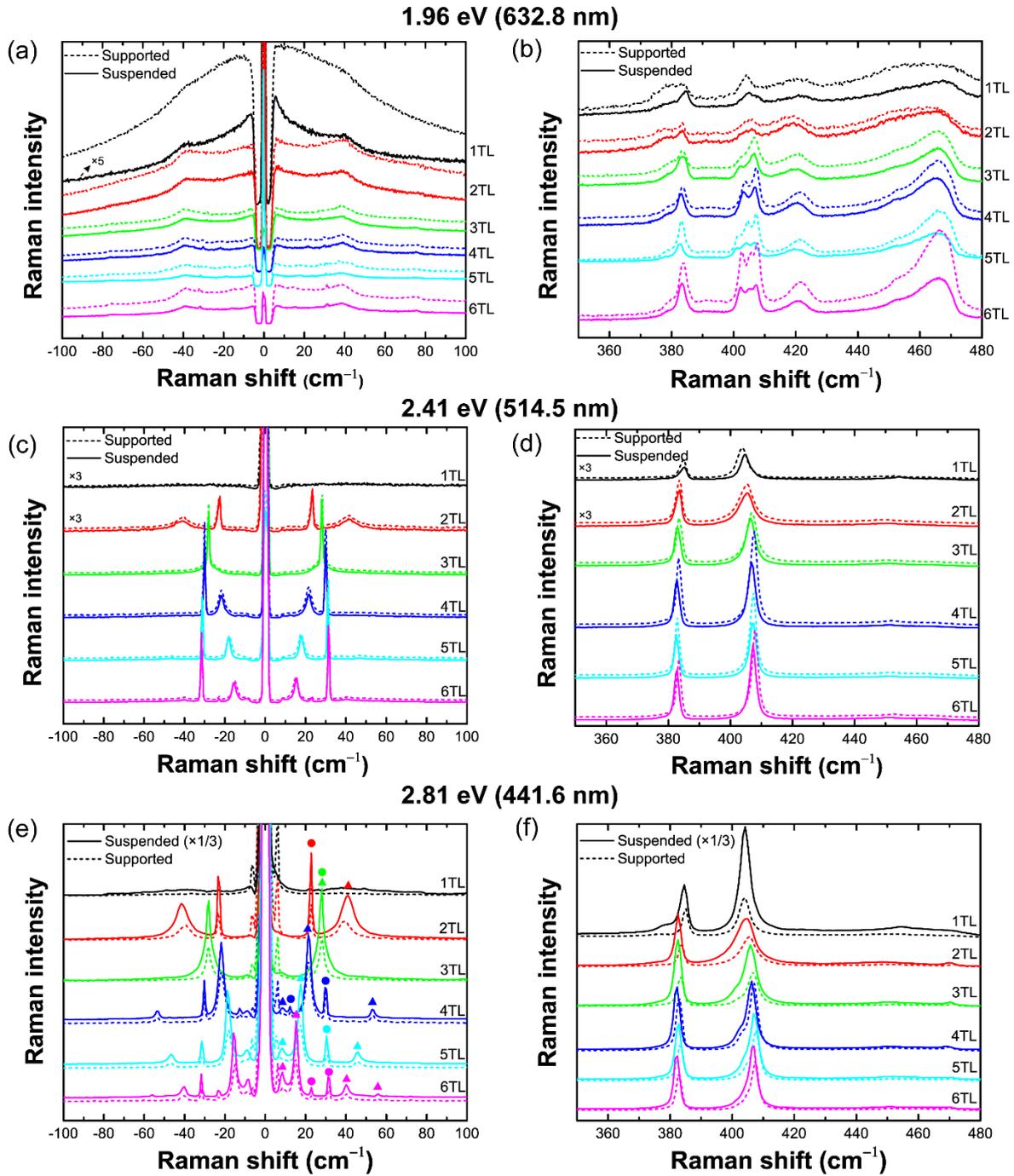

**Figure 4** Raman spectra of suspended (solid lines) and supported (dashed lines) MoS$_2$ (a,c,e) low-frequency region and (b,d,e) main modes measured with (a,b) 1.96, (c,d) 2.41, and (e,f) 2.81 eV excitation energy. The shear (●) and breathing (▲) modes are indicated.